\documentstyle[prl,aps,multicol]{revtex}

\begin{document}

\draft
\widetext

\title{Angular Dependences of Third Harmonic
  Generation from Microdroplets}

\author{J. Kasparian$^{(a)}$, B. Kr\"amer$^{(b)}$, 
J. P. Dewitz$^{(c)}$\cite{email}, S. Vajda$^{(b)}$, 
P. Rairoux$^{(b)}$\cite{awi}, B. Vezin$^{(a)}$, V. Boutou$^{(a)}$,
T. Leisner$^{(b)}$, W.~H\"ubner$^{(c)}$, J.~P.~Wolf$^{(a)}$,
L.~W\"oste$^{(b)}$, K.~H.~Bennemann$^{(c)}$}

\address{(a) LASIM (UMR 5579), Universit\'e Claude Bernard Lyon 1, 43 bd 
du 
11
  novembre, 69622 Villeurbanne Cedex France} 

\address{(b) Institut f\"ur Experimentalphysik, Freie Universit\"at
  Berlin, Arnimallee 14, 14195 Berlin, Germany}

\address{(c) Institut f\"ur Theoretische Physik, Freie Universit\"at
  Berlin, Arnimallee 14, 14195 Berlin, Germany.}

\date{\today}

\maketitle

\begin{abstract}
We present experimental and theoretical results for the angular
dependence of third harmonic generation (THG) of water droplets in the
micrometer range (size parameter $62<ka<248$). The THG signal in $p$-
and $s$-polarization obtained with ultrashort laser pulses is compared 
with a recently developed nonlinear
extension of classical Mie theory including multipoles of order
$l\leq250$. Both theory and experiment yield over a wide
range of size parameters remarkably stable intensity maxima close to
the forward 
and backward direction at ``magic angles''. In contrast to linear
Mie scattering, both are of comparable intensity.
\end{abstract}
\begin{multicols}{2}   
\narrowtext

Studies of optical microcavities have received increasing interest both from 
fundamental~\cite{kni} as well as applied research~\cite{arn,gra}.
The quality factor of microcavities formed by liquid droplets can become
as large as $10^{6}$. This leads to very high internal electromagnetic fields 
under resonant excitation. Accordingly, nonlinear optical effects
including stimulated Raman scattering (SRS), third order sum frequency
generation (TOSF) and in particular third harmonic generation (THG) in
microparticles are enhanced by several orders of magnitude compared to
flat surfaces and bulk media~\cite{qia,sta}.

These enhanced effects, in turn, make
nonlinear optics a promising spectroscopic probe for the 
characterization of microparticles. With respect to applications
in this area a lot of interest is focused on 
determining size, shape, and refractive index from the
measured intensities. With the help of the
recently developed lasers generating ultra-short pulses of high
intensities, higher harmonic generation can thus act as an additional
source of information.

Optical TOSF in particular THG in droplets was first reported by
Chang and coworkers in many organic and inorganic
liquids~\cite{ack,lea}. Theoretical models for TOSF and angle
integrated SHG in microdroplets have been proposed by several
authors~\cite{hil}. The angular dependence of SHG and THG by particles
with 
size parameter $ka\leq 5$ ($ka=2\pi a/\lambda$
 , $a$ being the radius of the particle) has recently been
 investigated by Dewitz {\em et al.} using a
 nonlinear extension of classical Mie theory~\cite{dew1}.

Here we report the first observation of the angular
dependence of THG generated by micrometer-sized water droplets
ranging from 8 to 32~$\mu$m ($62<ka<248$), as well as new
calculated angular distributions for droplets with radius up
to about 10.5~$\mu$m ($ka\approx 81$). Theory and experiment 
yield in fair agreement the following three remarkable features:

-The THG scattering profile of droplets having a radius of $a\gtrsim 8\;\mu$m 
is dominated by only a few intensity maxima at
magic angles close to the forward and backward directions. 

-The intensity 
and the angular distribution of the THG signal depends only weakly on 
the particle size and the central wavelength of the laser. 

-Theory and 
experiment find comparable intensity close to the forward and backward
direction. 

Each of these findings is in strong contrast to linear Mie scattering and is 
of considerable relevance for the application of nonlinear optical 
spectroscopy
to micro-particle sizing and detection. 

Our theoretical model for higher harmonic generation by
microdroplets is similar to the formalism of linear Mie
theory~\cite{mie}. Therein the electrodynamic
response of spherical particles is described by matching the incident,
internal and scattered field with the help of the boundary conditions
of their parallel components. The expansion coefficients
of the scattered field can then be expressed in terms of the
coefficients of the incident field. On this basis \"Ostling {\em et
  al.}~\cite{sta} developed a model for angle-integrated higher
harmonic 
generation by 
microparticles. This theory has been extended to angular dependences by
three of  the present authors~\cite{dew1}. The main 
idea of the approach is that in a first step the incident
field generates a charge $\sigma=\sum^\infty_{n=1}\sigma^{(n)}$, where
$\sigma^{(n)}$ oscillates with frequency $n\omega$. This picture is
motivated by the anharmonic oscillator model~\cite{dew1}. In a second
step the nonlinear components $\sigma^{(n)}$ radiate the $n$-th
harmonic field. In the central approximation of the theory
$\sigma^{(n)}$ is set equal to $[\sigma^{(1)}]^n$, where
$\sigma^{(1)}$ is equal to the radial component of the polarization
  ${\bf P}^{(1)}$ known from linear Mie theory~\cite{sta}. The $n$-th
  power of
 $\sigma^{(1)}$ can easily be
  obtained by expanding $\sigma^{(1)}$ in
terms of spherical harmonics taking into account the appropriate selection
rules for the angular momentum and keeping only terms varying with
$e^{{\rm i}n\omega t}$. Using $\sigma^{(n)}$ in combination with the
boundary conditions ${\bf
  n}\cdot[{\bf D}^{(n)}_{\rm out}-{\bf D}^{(n)}_{\rm
  in}]=4\pi\sigma^{(n)}$ for the radial components and $
{\bf n}\times[{\bf E}^{(n)}_{\rm out}-{\bf E}^{(n)}_{\rm in}]=0$ for
the parallel components and the appropriate expansion of the fields in
vector spherical harmonics~\cite{jac}, one determines then the
radiated intensities from the radial part of the Poynting vector
$|{\bf n}\cdot({\bf E}\times{\bf H})|$. Here the subscripts ``in'' and ``out''
denote the fields inside and outside the droplet~\cite{add1}. 

In THG 
multipoles of order $l\leq3\cdot ka+\delta$
($\delta\approx 20$ according to the required precision) must be
kept in the multipole expansions in order to get a fully converged
series expansion for the fields. Although conceptionally
straightforward it is numerically very
time-consuming to obtain higher-order multipole contributions.
Consequently our previous calculations were restricted to size
parameters $ka\leq 5$. For {\em small} spherical droplets in this size range
the theoretical treatment showed for SHG and THG an increasing number of maxima
which where distributed over the whole angular profile. They result from an
increased number of multipoles. The angles of the intensity maxima
exhibited strong variations upon small changes of the droplet size
parameter $ka$. Moreover, the enhancement of the intensities scattered
in the forward direction -well known from linear Mie
scattering~\cite{borwol,bohhuf}-, was amplified in SHG and THG 
dominating the angular profile. In the present paper we focus on larger
particles ($ka > 62$) obtaining quite distinct results.

The experimental setup is given in Fig.~\ref{fig2}. Monodisperse water 
droplets were generated at a repetition rate of 1~kHz by a
piezo-driven nozzle. Depending on the operation condition of the
nozzle, droplets in the size range between $8~\mu$m and $32~\mu$m could 
be produced. The droplet generation was synchronized to a Ti:Sa
femtosecond laser system, which produces ultrashort light pulses
(80~fs) with pulse energies up to $400~\mu$J at a central wavelength
of 820 nm. The scattered light was collected with a quartz fiber
mounted onto a stepper motor driven goniometer,then spectrally
dispersed into a UV spectrometer and detected by a photomultiplier
tube. Polarizing beamsplitters could be inserted in front of the fiber
to allow polarisation sensitive measurements. The setup allowed to
analyze the radiation emitted from the droplet in an angular range
between $18^\circ < \theta < 160^\circ$. The angular resolution was $1^\circ$.
The shot to shot variation of the particle size was controlled
independently by the analysis of the Mie scattering patterns obtained
with a He:Ne laser and by direct observation of the droplets under a
light microscope; it was found to be below 2~$\%$. The stability of the 
droplet generation and the temporal and spatial overlap of the Ti:Sa
laser beam and droplet stream was continuously monitored by observing
the scattered laser light with a microscope which was equipped with a
CCD camera readout. At higher fluences of the Ti:Sa laser, the droplets 
disintegrated into a cloud of small droplets with a diameter of about
$1~\mu$m.
 
Since the optical 
response of the droplets to the ultrashort laserpulses is much faster
than their fragmentation, the light scattering signal 
was nevertheless detected. It was checked experimentally that
background signal from remaining small aerosol droplets could be
neglected compared to the initial droplet signal. In order to
determine the dominant channel of the nonlinear optical response we
performed spectroscopic measurements. We find as the most significant
feature a large third harmonic peak at 273~nm. It showed a cubic
dependence of the incident laser fluence, as
expected for a third order process. 
The SHG peak at 410~nm is rather weak. This is an indication for the good
spherical shape of the droplets for which SHG is forbidden by
symmetry. The THG light was $s$- or $p$-polarized as the incident light, 
the amount of the cross polarization was
only 2$\%$. Moreover, no stimlated Raman scattering was observed. This is 
comprehensible with respect to the short laser pulses used and demonstrates
that THG is indeed the dominant nonlinear
response channel. 

In Fig.~\ref{fig3} the experimental results for the angular
dependence of THG of water droplets having a radius of $a=10.5\;\mu$m
($ka\simeq 81$) are shown for angles between 18$^\circ$ and 160$^\circ$ in 
the $\varphi=0^\circ$ plane. Two pronounced intensity maxima
appear around 26$^\circ$ and 155$^\circ$, whereas almost no
intensity is detected at angles between 50$^\circ$ and 130$^\circ$. 
The result is in clear contrast to the angular dependence expected
for linear Mie scattering from droplets with $ka\approx 81$. Instead
of a very complex structure with several maxima of varying angle and
magnitude
only one broad maximum appears close to the forward and backward
directions. 

Moreover we find that the angular dependence is remarkably stable upon 
variation of the droplet size: the scattering profile is reproduced
for particles from $a=8\;\mu$m to 32~$\mu$m. Only a small shift of the
``magic angle''
of maximum intensity occurs from 26$^\circ$ to 30$^\circ$. Thus the
angular 
dependence does not
only show a strong reduction of complexity but also a {\em stabilization of
the angles} of intensity maxima over a wide range of droplet sizes,
although 
the number
of contributing multipoles varies strongly with $ka$.

Most importantly we find that the magnitude of the intensity maxima
close to the forward and backward direction is of the same
order. This is in sharp contrast to linear Mie scattering where
forward scattering is strongly enhanced.

In Fig.~\ref{fig4} we show the theoretical results for the
angular dependence of the THG intensities of a water droplet with
radius $a=10.5\;\mu$m ($ka\approx 81$) in the full $\theta$-range from
0$^\circ$ to 
180$^\circ$,
in the inset the corresponding polar plot is displayed. Dominating
maxima 
appear close to the forward and backward direction
at angles $\theta\approx 10^\circ$ (where no experimental data could
be recorded) and $\theta\approx 158^\circ$. Those maxima show unexpected 
stability upon droplet size
variation. The stability of intensity maxima in THG strongly contrasts
with linear Mie scattering, where theory yields a strong change of the
scattering profile upon variation of $ka$ in a small range around
$ka\approx 81$~\cite{jpd}. Furthermore the magnitudes of the intensity
maxima close to the forward and backward directions are of same order
in THG as it is observed in experiments (see Fig.~\ref{fig3}). This is again in
clear contrast to linear Mie scattering for which the calculated ratio
of forward to
backward scattering for water droplets with $ka\approx 81$ is around 
6700~\cite{jpd}.

The stabilization of the angular scattering profile in THG as a
function of $ka$ results from dominating resonances occurring within
the laser bandwidth. These sharp resonance peaks are enhanced by six
orders of magnitude as compared to the background radiation. The dramatic
enhancement results from the nonlinearity in THG, which amplifies the
field effects. Remarkably, our calculations yield that every resonance
displays the same angular pattern. Experimentally these resonances are
always excited due to the finite laser bandwidth of $\pm5$~nm resulting
from the femtosecond pulses and the particle size fluctuation of
$\pm 2$\%. In our case these effects have been fully taken into account in
the calculations. In the experimental $ka$-range ($ka=81 \pm 2$)
four resonances are excited.

In backward direction the 
position of the two peaks with the largest angles (156$^\circ$, 160$^\circ$) 
are found by both theory and experiment. For decreasing angle the
agreement fades since the theory shows a more complicated structure
and thus a stronger oscillation of the intensities. The damping of the
experimental structure results from the finite
angular resolution which is of the order of 1$^\circ$. For intensities
less than 130$^\circ$ both theory and experiment show no structure.

In the forward direction theory and experiment agree in a less
satisfactory way. 
Calculations predict a strong maximum at an angle of 10$^\circ$. 
This, however, is not observable with our existing equipment.
Experimentally a maximum is observed at 
26$^\circ$.
Work is in progress to investigate the
angular range also below 18$^\circ$. 

In conclusion, the angular dependence of THG of water
droplets with sizes from 8~$\mu$m up to 32~$\mu$m are studied
experimentally and for sizes around 9 and 10~$\mu$m theoretically. In
good 
agreement both investigations show a reduced complexity of the scattering
profile and a 
stabilization of the angles of maximum
intensities 
close to the
forward and backward direction. Moreover the magnitude of the intensities
in the forward and backward direction are of the same order of
magnitude. These results are in remarkable contrast to the
characteristics of linear Mie scattering in this size range. 

The angles of the maxima near the backward direction show good
quantitative agreement between theory and experiment. It was shown
that the stabilization is an effect of
the excitation of sharp resonance peaks with stable angular dependence
independent of the size of the droplet.

This stabilization can be considered in analogy to rainbow scattering
which occurs in linear optics for large droplet sizes and reflects the
onset of a transition from Mie scattering to geometrical optics. The
latter is valid for a plane surface resulting from the limit of
droplets with infinite radius. Due to the incorporation of higher
multipoles in THG this limit is reached for smaller droplets than in
linear optics.

Thus THG opens up new possibilities to characterize 
particles in the Mie-size range between 
Rayleigh-scattering and
geometrical optics. Especially the strongly enhanced backward scattering
yields new perspectives for the remote sensing
of the atmosphere.

J. P. Dewitz and W. H\"ubner gratefully acknowledge financial support
through Deutsche Forschungsgemeinschaft Sfb 290.
The autohrs of LASIM also acknowledge the DRET for actively supporting
this project.


\end{multicols}
\begin{figure}
\caption{Experimental setup} 
  \label{fig2}
  \caption{Measured spectra of THG intensity of a water droplet with
    $a=10.5\mu$m as a function of $\theta$ for parallel
    polarization. The forward and backward scattering is recorded
    separately as indicated by the dashed line. Thus the absolute
    intensity of both scans cannot be compared exactly.}
\label{fig3}
\caption{Theoretical spectra of the THG intensity of a water droplet
    with $a=10.5\;\mu$m in arbitrary units, represented linearly
    in the main part and as a polar plot in the inset. Parallel polarization is
    considered.}
\label{fig4}
\end{figure}

\begin{references}
\bibitem{add1}As a consequence
of the approximation $\sigma^{(n)}\equiv [\sigma^{(1)}]^n$ the nonlinear
susceptibility $\chi^{(n)}$ is taken to be constant and set equal to
unity, 
thus the absolute value of the radiated intensities
cannot be calculated. However, the theory still yields the full
information 
about the
angular dependence of the intensities and their variation with size,
shape, and refractive index, since all interferences of the
contributing 
multipoles are
obtained.
\bibitem{add2}Note however, that for ideally transparent objects interferece
effects similar to the shape resonances discussed in our paper cannot
be neglected.
\end{references}
\end{document}